\documentclass[11pt]{article}

\newcommand{\pa}{\partial}

\newcommand{\lb}{\left[}
\newcommand{\rb}{\right]}
\newcommand{\be}{\begin{equation}}
\newcommand{\ee}{\end{equation}}

\newcommand{\z}{\zeta}
\newcommand{\ra}{\rightarrow}

\usepackage{amsmath,amsthm}
\usepackage{amssymb,latexsym}
\usepackage[mathscr]{eucal}
\usepackage{graphics}
\usepackage{setspace}
\usepackage{array}
\numberwithin{equation}{section}

\begin{document}

\title{Resonant solitons from the $3\times 3$ operator}

\author{D. J. Kaup$^{1*}$ and Robert A. Van Gorder$^1$ \\  \small $^1$ Department of Mathematics, P.O. Box 161364, University of Central Florida, Orlando, FL  32816-1364 USA\\
\small $^*$ Corresponding author. Email: kaup@mail.ucf.edu}
\date{\today}
\maketitle

\centerline{ \noindent\Large\textbf{Abstract}}
We study and detail the features of the resonant soliton of the $3\times 3$ operator. The scattering data of this operator contains four transmission coefficients, two in each half complex $\zeta$-plane, where $\zeta$ is the spectral parameter. When the potential matrix has anti-hermitian symmetry ($Q_{\alpha,\beta}=-Q^*_{\beta,\alpha}$), each of the two transmission coefficients in the lower half plane become equal to the complex conjugates of one of the two in the upper half plane, leaving only two independent transmission coefficients, which we take to be those in the upper half plane. The bound state scattering data for this operator consists of the zeros of these two transmission coefficients (bound state eigenvalues) and a normalization coefficient associated with each eigenvalue.  With two transmission coefficients, there is a wider variety of possible soliton solutions than in the Zakharov-Shabat (ZS) $2\times 2$ case. First, the eigenvalues could belong to only one transmission coefficient, with the other transmission coefficient having none. In this case the soliton solution will only exist in one of the two conjugate pairs of components of $Q$ lying next to the diagonal, with all other components of $Q$ vanishing.  These soliton solutions will be equivalent to those in the ZS case, up to scaling factors. If we would distribute the eigenvalues among both transmission coefficients so that each transmission coefficient would have at least one eigenvalue, the soliton structure becomes more complex. In this case, any one eigenvalue will be found to generally make contributions to all non-diagonal components of $Q$. Of particular interest, inside this class is a special class of solitons which exist only when these two transmission coefficients have exactly equal eigenvalues. Equality of eigenvalues from different transmission coefficients give rise to ``resonant solitons".  This latter class is the case which we treat here.  We find that there are two different states for the resonant soliton solution, both of which arise from a bifurcation.  The bifurcation is shown to arise from the algebraic structure of the $3\times 3$ scattering matrix. We also extend the linear dispersion relations (the solution of the inverse scattering problem) to the case when resonant solitons are present.  We further show that the same result could be obtained by taking the limit of the eigenvalues in each transmission coefficient approaching the other. We then study the general soliton solution under anti-hermitian symmetry when there is only one eigenvalue in each transmission coefficient, with each unequal to the other.  We then detail the asymptotics of this solution and then the same when the eigenvalues are exactly equal, showing that the latter contains the well known parametric interactions of "up-conversion" and "down-conversion". Lastly, we explain how this equality of eigenvalues in different transmission coefficients can be seen to be a nonlinear resonance condition. 
\\
\\


\section{Introduction}

The purpose of this paper is to detail the features of the resonant soliton solutions of the nondegenerate $3\times 3$ eigenvalue problem
\be\label{1.1}
\pa_x V - i\z J\cdot V = - Q\cdot V\,,\quad
J = \lb \begin{matrix} J_1 & 0 & 0 \\ 0 & J_2 & 0 \\ 0 & 0 & J_3 \end{matrix} \rb\,, \quad
Q = \lb \begin{matrix} 0 & Q_{12} & Q_{13} \\ Q_{21} & 0 & Q_{23} \\ Q_{31} & Q_{32} & 0 \end{matrix} \rb
\ee
on the interval $-\infty < x < +\infty$.  We assume $J_1>J_2>J_3$ and that $\text{Tr}(J)=0$.  $Q (x)$ is a potential matrix which vanishes like $Q (x\ra \pm \infty) = o (1/x)$ for large $x$. The matrix $V (x)$ is a $3 \times 3$ solution matrix which contains Jost solutions as its columns. This eigenvalue problem provides the Inverse Scattering Transform (IST) for the three-wave resonant interaction (3WRI) \cite{ZM,Pub020,ZM2,Pub041}. Here, we generally assume no symmetry for the matrix $Q$ and will take the six components of $Q$ to be generally independent and uniquely different. 

The Lax pair for the 3WRI nonlinear system was first presented in 1973 \cite{ZM} along with the soliton solutions for the explosive and decay cases. Here the components of $Q$ are the envelopes for interacting waves.  Here for the first time, it was demonstrated that in the explosive case, there was a nonlinear instability whereby a singularity with infinite amplitude would eventually develop in the solution. In 1975, approximately at the same time, two manuscripts were submitted \cite{Pub020, ZM2} giving more details on this system and its solutions.  The work by Kaup \cite{Pub020} presented a derivation of the IST for this system, giving a set of Marchenko equations and described the solutions from the point of view of the Marchenko equations.  In the work by Zakharov and Manakov \cite{ZM2}, they constructed various soliton solutions and described the interactions from the point of view of these soliton solutions and the quasi-classical approximation. Each work made use of separable initial data for constructing the general scattering matrix.  The final solution and the entire interaction could then be described in terms of the initial soliton and radiation (continuous spectra) content of each wave. In 1979, Kaup, Reiman and Bers \cite{Pub041} publish a review of the 3WRI wherein more detailed aspects of these solutions were given along with comparisons and validations from numerical solutions. In a related work \cite{Reiman}, Reiman extended these results to the case where the medium could contain spatial inhomogeneities and studied how they differed from the homogeneous solutions and how they modified those results.  Interestingly, Reiman showed that the inhomogeneous case of the 3WRI was also an integrable system solvable by the same IST of the homogeneous 3WRI. Higher order 3WRI soliton solutions, particularly those corresponding to multiple zeros in the transmission coefficients, have recently been given in Ref. \cite{Yang03}.  A perturbation treatment of the decay instability was briefly treated in \cite{Pub041}.  More recently, a full treatment of the perturbations of \eqref{1.1} and the universal covering set for the squared eigenfunctions and their adjoints has been given in Ref. \cite{Pub249}, including the closure and the orthogonality relations. The general nature of the soliton solutions of the $n\times n$ generalization of \eqref{1.1} has been described in \cite{ZMNP}. These soliton solutions have also been classified in Ref. \cite{Pub230} according to the group properties of the eigenvalue problem, while a recent overview of the $n\times n$ IST and its soliton structure is found in Ref. \cite{Pub250}. 


For the simple $2\times 2$ case \cite{ZS72, AKNS74}, one has only one type of soliton, generally called an ``NLS (Nonlinear Schr\"odinger)" soliton.  Due to the group properties of the Zakharov-Shabat (ZS) eigenvalue problem, the NLS soliton is classified as a $sl(2)$ soliton \cite{Pub230}.  The soliton solutions that arise in the $3\times 3$ case are classified as $sl(3)$ solitons.  The general soliton solutions for three of the $sl(3)$ symmetries have been given in \cite{ZM2, Pub020, Pub041}. 

In the 3WRI, the upper triangular components of $Q$ correspond to three different waves which can interact resonantly. (The lower triangular components of $Q$ are $\pm$ the complex conjugates of those in the upper triangle.) There are two general types of $sl(3)$ soliton solutions. First, depending on the symmetry, any one of the three waves could contain one or several NLS solitons, provided all the solitons were found in only one wave.  This is a trivial example of the $sl(3)$ solitons wherein one has only one nonzero wave, not three. Thus no interactions occur. These $sl(3)$ solitons are identical to $sl(2)$ $N$-soliton solutions.  This is expected since $sl(2)$ is a subgroup of $sl(3)$.  Thus there would be three different representations of the $sl(2)$ subgoup, one for each wave. The second general type is the nontrivial case, wherein all three waves are nonzero. For these $sl(3)$ solitons, they interact with their energy being exchanged back and forth between the three waves in a regular fashion. However the asymptotics of these solutions show that in general, no solitons are exchanged between the three waves. 

In the general $2\times 2$ eigenvalue problem where the two off-diagonal components of $Q$ are independent and unrelated, there are two transmission coefficients, one in each complex half-plane, the zeros of which, along with a normalization coefficient associated each zero, determine the $sl(2)$ soliton structure.  In the case of the $3\times 3$ eigenvalue problem, one has a total of four transmission coefficients, two in each complex half-plane.  The zeros of these, along with a normalization coefficient for each, determine the $sl(3)$ soliton structure. Amongst these structures lies one that is uniquely different from $sl(2)$ solitons, which is called a ``resonant soliton" \cite{Pub249, Pub250}. A resonant soliton is one where the two transmission coefficients in each half-plane have a common zero which gives a bifurcation in the general $sl(3)$ soliton case mentioned above. It is unique among the $sl(3)$ soliton solutions in that not only is energy exchanged between the waves, but also entire solitons can be exchanged. This feature of the 3WRI, whereby one can transfer packets of energy from a wave of one frequency to another wave of a different frequency, and/or back again, has long been suggested as a basis for possible designs of optical logical elements.  


Recently this feature has generated further interest for the same purposes in a different setting. Here one utilizes a nonzero background wherein one takes a combination of bright and dark solitons.  Degasperis, et al \cite{App04, App02, App06} have found exact solutions of this 3WRI when one wave is taken to be on an asymptotically constant nonzero background, with the other two being localized pulses.  Solutions exist where the three interact and move with a common velocity (simultons). Such solutions were found to be stable when their velocity was greater than a critical value. Interactions of simultons with various localized pulses were found to give rise to the excitation (decay) of stable (unstable) simultons by means of the absorption (emission) of the energy carried by a localized pulse. The speed of these solitons could be continuously varied by means of adjusting the energies of the two bright pulses.

Here we shall only treat the bright soliton case. The major purpose of this paper is to detail those features of the scattering data which allow this soliton resonance phenomenon to occur. Secondary will be to describe how these features will display themselves in the potential matrix, $Q$.   


As is known, physical applications of the IST typically require the potential $Q$ to have certain symmetries. In the $2\times 2$ case on the infinite interval, the symmetry for most physical applications are found in one of the two cases $Q_{21} = r = \pm q^* = \pm Q^*_{12}$ where $ ^*$ indicates the complex conjugate. If $Q(x\rightarrow\pm\infty)\rightarrow 0$, then solitons can only occur in $r=-q^*$ case \cite{ZS72,AKNS74}.  If one allows for an asymptotically constant nonzero background, then in the $r=+q^*$ case one can have ``dark solitons" \cite{ZS73}.  If the solution is to exist only on a semi-infinite or finite interval, {\it virtual} solitons \cite{Pub195,Pub202} for either symmetry can be found. 

For the $3\times 3$ case, one has several other possible symmetries, three of which are found in the 3WRI \cite{ZM2, Pub020, Pub041}. Here we shall only consider the general case of no symmetry or the symmetry $Q_{\alpha, \beta} = - Q^*_{\beta, \alpha}$, which we shall call the anti-hermitian symmetry. This latter symmetry in the $sl(3)$ case corresponds to the soliton decay case of the 3WRI and in the $sl(2)$ case, to $r=-q^*$.   

In Section 2, we will briefly outline the results from the direct and the inverse scattering problems for Eq. \eqref{1.1}.  For the case of no symmetry, we give the linear dispersion relations (LDRs) from which one defines the scattering data.  For the anti-hermitian symmetry, the bound state scattering data consists of the zeros of two different transmission coefficients, with a normalization coefficient associated with each zero.  From this structure, as discussed earlier, there are two general types of $sl(3)$ soliton solutions. First, zeros could be in only one transmission coefficient, with the other transmission coefficient having no zeros. In this case, one has only an $sl(2)$ $N$-soliton solution which would only exist in one of the two conjugate pairs of components in $Q$ adjacent to the diagonal. For the second type, the zeros would be distributed between both transmission coefficients. This soliton solution would make contributions to all three conjugate pairs of $Q$. To specify these $sl(3)$ soliton solutions, we use the notation $(M,N)$-soliton solution, where $M$ will be the number of zeros of the first transmission coefficient and $N$ will be the number of zeros in the second transmission coefficient. We also give and discuss the general $sl(3)$ $(1,1)$-soliton solution.  When these two eigenvalues are equal, we have a resonant soliton. 

In Section 3 we discuss the general $sl(3)$ soliton solution when an arbitrary number of resonant solitons are present.  We find that there are two distinctly different structures for a resonant soliton, due to a bifurcation in the scattering matrix. These structures are obtained for each branch of the bifurcation, for the case of compact support. Also the LDRs are extended to include resonant solitons. We then briefly discuss how to handle the non-compact support case and how it will give the same structures for resonant solitons. In Section 4, under anti-hermitian symmetry, we obtain the resonant $sl(3)$ $(1,1)$-soliton solution from the extended LDRs. Within this solution, we find the third representation of the $sl(2)$ subgroup and delineate its scattering data. In Section 5, we briefly discuss how the resonate soliton can also be obtained by taking the limit of the two zeros approaching each other in the general $sl(3)$ $(1,1)$-soliton solution.  The same can be also be done to obtain the extended LDRs. A summary of the results and conclusions is given in the last section. 

\section{The Direct and Inverse Scattering Problems}

Here we highlight the direct and inverse scattering problems for the general $3\times 3$ operator given in \cite{Pub249}, which followed the method used in Refs. \cite{ZM,Pub020,ZM2,Pub041}.  This method allows one to obtain a linearly independent set of the LDRs, from which one could reconstruct the potentials.  Later, we shall employ these LDRs to obtain the general (1,1)-soliton solution for the $3\times 3$ operator.

\subsection{The Direct Scattering Problem}

In the direct scattering problem, one addresses the solutions of the eigenvalue problem, what their analytical properties are, the adjoint solutions and their properties, what is the scattering matrix and its properties, what are the features of the bound states, if any, and what are the fundamental analytical solutions and their adjoints, etc. Each of these topics shall only be defined and discussed to the extent necessary for the construction of soliton solutions. Further details will be found in the above references, \cite{ZM,Pub020,ZM2,Pub041,Pub249}.

The relevant matrix form of the linear eigenvalue problem associated with the 3WRI is given by \eqref{1.1}.
We assume no symmetry on $Q$ and take the six components of $Q$ to be independent (all the diagonal components are zero). We shall take the diagonal elements of $J$ to be real and to satisfy $J_1>J_2>J_3$. $Q (x)$ is a potential matrix with zero diagonal entries and vanishing as $|x| \rightarrow \infty$.  The matrix $V (x)$ is a $3 \times 3$ solution matrix which contains the Jost solutions as its columns. For \eqref{1.1}, we shall define the Jost solutions by their asymptotics as $x\rightarrow \pm \infty$.  For $\z$ real, there are two standard sets which are
\be\label{2.4}
\Phi (x \ra -\infty) \ra e^{i \z Jx}, \quad ~\Psi (x \ra +\infty) = e^{i \z Jx}\,.
\ee
Since these two matrices are linearly dependent on the other, we have
\be\label{2.5}
\Phi = \Psi \cdot S, \quad ~ \Psi = \Phi \cdot R,
\ee
where due to the Wronskian relation,
\be\label{2.5a}
\text{det} S = 1, \quad ~ \text{det} R = 1,
\ee
%
while the inverse relation is 
\be \label{RasS}
R = \lb \begin{matrix}
S_{33}S_{22}-S_{32}S_{23} & S_{32}S_{13}-S_{33}S_{12} & S_{23}S_{12}-S_{22}S_{13} \\
S_{31}S_{23}-S_{33}S_{21} & S_{33}S_{11}-S_{31}S_{13} & S_{21}S_{13}-S_{23}S_{11} \\
S_{32}S_{21}-S_{31}S_{22} & S_{31}S_{12}-S_{32}S_{11} & S_{22}S_{11}-S_{21}S_{12}
\end{matrix} \rb .
\ee

We will use $\pm$ superscripts to indicate the regions of analyticity ($+$ for the UHP and $-$ for the LHP) where UHP stand for the upper half complex $\zeta$-plane and LHP for the lower half. Subscripts will used to indicate the components of a matrix quantity. The individual Jost solutions are the columns in the Jost solution matrix. For this system, we have
\be\label{defJost}
\Phi = [\phi^-_1, \phi _2, \phi^+_3]\,, \quad ~\Psi = [\psi^+_1, \psi _2, \psi^-_3]\,, \quad
\ee
where those components without a $\pm$ superscript in general only exist on the real $\z$-axis. (Strictly speaking, it is not the Jost solution which is analytic in $\z$, uniformly for all $x$, but the various columns in matrix products such as $\Phi \cdot e^{-i\z Jx}$
. With this understood, we shall refer to the appropriate Jost solutions as being analytic in $\z$ if this product is so analytic.)
How to generally determine the analytical properties of the Jost solutions are detailed in the above references and textbooks such as Ref. \cite{Shabat, ZMNP}.

The analytical properties of the scattering matrices $S$ follow from the above and are
\be\label{S_a}
S = \lb
\begin{matrix}
S^-_{11} & S_{12} & S_{13} \\
S_{21} & S_{22} & S_{23} \\
S_{31} & S_{32} & S^+_{33}
\end{matrix} \rb\,, \quad
R = \lb
\begin{matrix}
R^+_{11} & R_{12} & R_{13} \\
R_{21} & R_{22} & R_{23} \\
R_{31} & R_{32} & R^-_{33}
\end{matrix} \rb\,.
\ee
In each region of analyticity, there are three linearly independent solutions of \eqref{1.1}.  For the real axis, those could be either $\Phi$ or $\Psi$, or a suitable mixture. For each half-plane, we already have two of these Jost functions. The third can be construct from a linear combination of the Jost functions as detailed by Shabat \cite{Shabat}.  Such solutions were constructed in \cite{Pub020}.  For inversion about $x=+\infty$, one can take our third independent function to be the meromorphic functions \cite{Pub249}
\be \label{defmu}
\mu_2^+\, =\, \psi _2  - {R_{12}\over R_{11}^+} \psi ^+_1 \,,\quad \quad
\mu_2^- \, =\,  \psi _2 - {R_{32}\over R_{33}^-}\psi ^-_3 \,.
\ee
We define the set of $\mu$ solutions, $\mu^+$ and $\mu^-$, as
\be\label{3.5x}
\mu^+ = [\psi^+_1, \mu_2^+, \phi^+_3]\,, \quad ~\mu^- = [\phi^-_1, \mu_2^-, \psi^-_3]\,, 
\ee

From these solutions, one may construct the ``fundamental meromorphic solutions" (FMS) which will be designated by $\Theta$. They are defined by
\be\label{2.14}
\Theta^\pm = \mu^\pm\cdot e^{-i\zeta J x} \,.
\ee
$\Theta^+$ provides us with a set of three linearly independent, meromorphic solutions in the UHP and existing on the real $\z$-axis, while $\Theta^-$ provides us with another similar set on the real $\z$-axis and in the LHP.

The remaining part of the direct scattering problem is to detail the asymptotics of the Jost solutions as one approaches any essential singularity on the boundary of the region of analyticity. This provides a means for obtaining the potentials, given the Jost solutions. There is only one essential singularity at $\mid \z \mid = \infty$ in this problem. Taking $\z$ to be real, then $\Phi$ and $\Psi$ have a common asymptotic expansion which is
\be\label{2.23}
\Phi\,,\Psi = ( I_3 + i{\cal B}^{(1)}/\z + {\cal B}^{(2)}/\z^2 + \ldots) \cdot e^{-i\z J x}\quad \text{as}\,\,
|\z| \ra \infty\,.
\ee
In terms of $\Theta^\pm$, this leads to
\be\label{2.24}
\Theta^\pm(|\z| \ra \infty) = I_3 + i{\cal B}^{(1)}/\z + O(1/\z^2) \,,
\ee
in the appropriate half-plane.
One finds that when no pairs of the components of $J$ are equal, then ${\cal B}^{(1)}$ can be given by
$[ {\cal B}^{(1)}, J ] = Q$, whose solution is
\be\label{2.26}
{\cal B}^{(1)} =  \lb \begin{matrix} X & \dfrac{Q_{12}}{J_1-J_2} & \dfrac{Q_{13}}{J_1-J_3} \\
\\ \dfrac{Q_{21}}{J_2-J_1} & X & \dfrac{Q_{23}}{J_2-J_3} \\  \\
\dfrac{Q_{31}}{J_3-J_1} & \dfrac{Q_{32}}{J_3-J_2} & X
   \end{matrix} \rb\,,
\ee
where the $X$'s are generally integrals over quadratic products of the components of the potential matrix, $Q$.

\subsection{The Inverse Scattering Problem}

On the real axis, which is the boundary between the two regions of analyticity, it follows that $\Theta^+$ and $\Theta^-$ will be related linearly.  Thus one may construct what are called ``linear dispersion relations" (LDRs), whereby one obtains a set of nonhomogeneous, linear, algebro-singular integral equations relating $\Theta^+$ in the UHP (and on the real axis) to $\Theta^-$ in the LHP (and on the real axis), and vice versa. As shown in \cite{Pub249}, one can reduce these six equations to a set of three linearly independent equations, from which one can define the scattering data and which also provides a solution for the inverse scattering problem, if it exists. (As is well known, sometimes potential components in $Q$ have singularities which are physical, as in the explosive case of the 3WRI \cite{ZM}.) An alternate derivation of the LDRs has been given in Ref. \cite{Pub250}, whereby one directly obtains a set of $3$ linearly independent LDRs for the $3\times 3$ case of \eqref{1.1}, which is obviously extendible to the problem of obtaining a set of $n$ linearly independent LDRs for the general $n\times n$ case of \eqref{1.1}. 

Taking the results from \cite{Pub249}, we have the four LDRs (of which only three are linearly independent on the real axis)
\be \label{2.13eqns}
\Theta^+_1(\z) = \lb \begin{matrix} 1 \\ 0 \\ 0 \end{matrix} \rb
   - {1\over 2\pi i} \int_{\cal R}{d\z' \over \z'-\z} \left[ {S_{21}\over S_{11}^-}
   \left( \Theta^-_2( \z^\prime ) E_{21}( \z^\prime )
   + {R_{32} \over R^-_{33}} \Theta_3^-( \z^\prime ) E_{31}( \z^\prime ) \right)
    + {S_{31}\over S_{11}^-}\Theta^-_3( \z^\prime ) E_{31}( \z^\prime )\right]
\ee \be\label{Thp1f}
   - \sum_{k = 1}^{N_{11}^-} \frac{1}{(\z_{11,k}^- - \z)}\, {S_{21,k} \over S_{11,k}^{-\prime}}\,
   \Theta^-_2(\z_{11,k}^-)E_{21}(\z_{11,k}^-)\,,
\ee
\be \label{2.14eqns}
\Theta^+_2(\z) = \lb \begin{matrix} 0 \\ 1 \\ 0 \end{matrix} \rb
 - {1\over 2\pi i} \int_{\cal R}{d\z' \over \z'-\z} \left[ {R_{12}\over R_{11}^+}\Theta_1^+( \z^\prime )
   E_{12}( \z^\prime ) - {R_{32}\over R_{33}^-}\Theta^-_3( \z^\prime ) E_{32}( \z^\prime )\right]
\ee\be\label{Thp2f}
   + \sum_{k=1}^{N^+_{11}} {1\over \z^+_{11,k}-\z}\,\, {R_{12,k}\over R^{+\prime}_{11,k}}\,
   \Theta^+_{1}({\z^+_{11,k}}) E_{12}(\z_{11,k}^+)\,
   +\sum_{k = 1}^{N_{33}^-} \frac{1}{(\z_{33,k}^- - \z)} \, {R_{32,k} \over R_{33,k}^{-\prime}}\,
   \Theta^-_3(\z_{33,k}^-) E_{32}(\z_{33,k}^-)\,,
\ee
and for $\Im{(\z)} \le 0$,
\be \label{2.15eqns}
\Theta^-_2(\z) = \lb \begin{matrix} 0 \\ 1 \\ 0 \end{matrix} \rb
 - {1\over 2\pi i} \int_{\cal R}{d\z' \over \z'-\z} \left[ {R_{12}\over R_{11}^+}\Theta_1^+( \z^\prime )
   E_{12}( \z^\prime ) - {R_{32}\over R_{33}^-}\Theta^-_3( \z^\prime ) E_{32}( \z^\prime )\right]\notag
\ee\be\label{Thm2f}
   + \sum_{k=1}^{N^+_{11}} {1\over \z^+_{11,k}-\z}\,\, {R_{12,k}\over R^{+\prime}_{11,k}}\,
   \Theta^+_{1}({\z^+_{11,k}}) E_{12}(\z_{11,k}^+)\,
   +\sum_{k = 1}^{N_{33}^-} \frac{1}{(\z_{33,k}^- - \z)} \, {R_{32,k} \over R_{33,k}^{-\prime}}\,
   \Theta^-_3(\z_{33,k}^-) E_{32}(\z_{33,k}^-)\,,
\ee
\be \label{2.16eqns}
\Theta_3^- = \lb \begin{matrix} 0 \\ 0 \\ 1 \end{matrix} \rb
   + {1\over 2\pi i} \int_{\cal R}{d\z' \over \z'-\z} \left[
   {S_{13}\over S_{33}^+}\Theta^+_1( \z^\prime ) E_{13}( \z^\prime ) +
   {S_{23}\over S_{33}^+}\left( \Theta_2^+( \z^\prime ) E_{23}( \z^\prime )
   + {R_{12}\over R_{11}^+} \Theta_1^+( \z^\prime ) E_{13}( \z^\prime )\right) \right]
\ee \be\label{Thm3f}
   -\sum_{k = 1}^{N_{33}^+} \frac{1}{(\z_{33,k}^+ - \z)} \, {S_{23,k} \over S_{33,k}^{+\prime}}\,
   \Theta^+_2(\z_{33,k}^+)  E_{23}(\z_{33,k}^+)\,.
\ee
As to the notation and quantities given above, $\mathcal{R}$ indicates that the path of the integral is along the real axis, $N_{11}^+$ $\left(N_{33}^+\right)$ is the number of zeros of $R_{11}^+(\z)$ $\left(S_{33}^+(\z)\right)$ in the UHP (assumed finite), $\z^+_{11,k}$ $\left(\z^+_{33,k}\right)$ is the $k$th zero of $R_{11}^+(\z)$ $\left(S_{33}^+(\z)\right)$ in the UHP, $N_{11}^-$ $\left(N_{33}^-\right)$ is the number of zeros of $S_{11}^-(\z)$ $\left(R_{33}^-(\z)\right)$ in the LHP (assumed finite), $\z_{11,k}^-$ $\left(\z^-_{33,k}\right)$ is the $k$th zero of $S_{11}^-(\z)$ $\left(R_{33}^-(\z)\right)$, $R^{+\prime}_{11,k}$ $\left(S^{+\prime}_{33,k}\right.$, $S^{-\prime}_{11,k}$, $\left. R^{-\prime}_{33,k}\right)$ is the value of $dR^+_{11}(\z)/d\z$ $\left(dS^+_{33}(\z)/d\z\right.$, $dS^-_{11}(\z)/d\z$, $\left. dR^-_{33}(\z)/d\z\right)$ at its $k$\underline{th} zero, and $R_{12,k}$, etc., in the case of compact support, are just the values of $R_{12}$, etc. at the appropriate zeros (and in the case of non-compact support, would just be some coefficients). Finally,
\be\label{defE}
E_{pq}(\z) = \exp{[i\z (J_p-J_q) x] }\,.
\ee
We observe that the integrands of the integrals in \eqref{Thp2f} and \eqref{Thm2f} are equal and oppose in sign while the discrete contributions in each are identical.  From their difference, taking the limit of the imaginary part of $\zeta$ vanishing, we have the relationship
\be\label{mid2}
\Theta_2^+ - \Theta_2^- = \frac{R_{32}}{R_{33}^-} \,\Theta^-_3\, E_{32}(\z) -
   \frac{R_{12}}{R_{11}^+}\, \Theta^+_1 \, E_{12}(\z) \,,\quad \Im{(\z)} = 0\,,
\ee
which according to \eqref{defmu}, is an identity.  Whence \eqref{Thp2f} and \eqref{Thm2f}, for $\Im{(\z)}=0$, are linearly dependent. However, since the other two equations require the values of $\Theta_2^\pm$ at the various poles, we need to retain both of these in the set of LDRs.  

This total set of nonhomogeneous, linear, algebro-singular integral equations is a minimal set of LDRs, from which one may reconstruct the Jost solutions, given the scattering data.

\subsection{The scattering data \label{ScatData}}

Once we have the LDRs in the above form, then it becomes possible to define the scattering data for this problem. In order to solve these equations, we first must specify the quantities:
\begin{itemize}
  \item the reflection coefficients $\sigma_{j1} = \dfrac{S_{j1}}{S^-_{11}}\quad (j=2,3),\quad \sigma_{j3} = \dfrac{S_{j3}}{S^+_{33}}\quad (j=1,2),\quad  \rho_{12} = \dfrac{R_{12}}{R^+_{11}}$ and $\rho_{32} = \dfrac{R_{32}}{R_{33}^-}$ on the real axis,
  \item the zeros of $R_{11}^+(\z)$ in the UHP $(\z^+_{11,k}; ~k = 1, 2, ..., N_{11}^+)$ and the values of $C_{12,k} = {R_{12,k}\over R^{+\prime}_{11,k}}$ at each such zero,
  \item the zeros of $S_{33}^+(\z)$ in the UHP $(\z^+_{33,k}; ~k = 1, 2, ..., N_{33}^+)$ and the values of $C_{23,k} = {S_{23,k} \over S_{33,k}^{+\prime}}$ at each such zero,
  \item the zeros of $R_{33}^-(\z)$ in the LHP $(\z^-_{33,k}; ~k = 1, 2, ..., N_{33}^-)$ and the values of $C_{32,k} = {R_{32,k} \over R_{33,k}^{-\prime}}$ at each such zero,
  \item the zeros of $S_{11}^-(\z)$ in the LHP $(\z^-_{11,k}; ~k = 1, 2, ..., N_{11}^-)$ and the values of $C_{21,k} = {S_{21,k} \over S_{11,k}^{-\prime}}$ at each such zero.
\end{itemize}
Note that we have six reflection coefficients but only four sets of eigenvalues and normalization coefficients. Under compact support, the four normalization coefficients would be the residues of four of the reflection coefficients at the appropriate eigenvalues. Thus there are two reflection coefficients, $\sigma_{13}$ and $\sigma_{31}$, which will not be associated with any normalization coefficients. In the linear limit, each reflection coefficient can be matched to the Fourier transform of one of the six components of $Q$. With only four normalization coefficients, we may specify the position and phase of solitons in only four of the potential components.  If solitons are to occur in the remaining two components, then their positions and phases cannot be given independently. Instead the positions and phases of solitons in these components must be determined by the positions and phases of the other four solitons.  It is this point that we want to further understand.  We shall return to this point later.  

Under the anti-hermitian symmetry, the adjoint solutions of \eqref{1.1} are linear combinations of the hermitian conjugate of the Jost solutions, $V$. As a consequence of this symmetry, $R_{\alpha, \beta}(\zeta) = S^*_{\beta, \alpha}(\zeta^*)$ which, on the real $\zeta$-axis, leads to $\sigma_{21} = \rho_{12}^*$ and $\rho_{32} = \sigma_{23}^*$. For the bound state data, $N_{11}^- = N_{11}^+$, $N_{33}^- = N_{33}^+$, $\z^-_{11,k} = \z^{+*}_{11,k}$, $\z^-_{33,k} = \z^{+*}_{33,k}$, $C_{21,k} = C_{12,k}^*$ and $C_{32,k} = C_{23,k}^*$.  Thus each quantity in the LHP is the Hermitian conjugate of another quantity in the UHP. This symmetry gives rise to nonsingular $sl(3)$ soliton structures which most closely match that of the familiar $sl(2)$ solitons where $r = -q^*$.  In the following subsection, we shall obtain the $(1,1)$-soliton solution under this symmetry, which is the simplest  nontrivial $sl(3)$-soliton solution. Using that, we will describe the resonant nature of this soliton solution. 

\subsection{The (1,1)-soliton solution}
The solution method for the $(M,N)$-soliton solution when there is no symmetry, is given in Appendix A.  We use those results to obtain the (1,1)-soliton solution for the anti-hermitian symmetry, provided that the two eigenvalues are distinct. To that end, take $N_{11}^+ = N_{11}^- = N_{33}^+ = N_{33}^- = 1$.  Since there is only one zero of each transmission coefficient, we can simplify the notation. We eliminate the triple subscripts on the eigenvalues, replacing them with only one subscript, replacing $\zeta^+_{11,1}$ with $\zeta^+_{1}$, $\zeta^-_{33,1}$ with $\zeta^-_{3}$, etc. We also drop the last subscript on the $C$'s since there is only one eigenvalue arising from each transmission coefficient. 

Turning to \eqref{2.24} and \eqref{2.26}, we may solve for the components of $Q$. First we impose the symmetry condition $Q_{\alpha \beta} = - Q^*_{\beta \alpha}$ which translates into the symmetry $R_{\alpha\beta}(\zeta) = S^*_{\beta\alpha}(\zeta^*)$ for the scattering matrices.  There are four fundamental quantities in the LDRs, on which the solutions depend.  These are given below and due to the assumed symmetry, can be reduced to only two $x$-dependent complex quantities.
\begin{eqnarray}
{\cal R}_{12} = &\frac{C_{12}}{2\eta_1}\,E_{12,+1} \, = &
   \, e^{i\gamma_{12}}\,e^{i\zeta^+_1(J_1 - J_2)(x-x_{12})} \, \label{R12} \\
{\cal S}_{21} = &\frac{C_{21}}{2\eta_1}\,E_{21,-1} \, = &{\cal R}_{12}^*\,\label{S21}\\
{\cal R}_{32} = &\frac{C_{32}}{ 2\eta_3}\,E_{32,-3} \, = &
   \, e^{-i\gamma_{23}}\,e^{-i\zeta^-_3(J_2 - J_3)(x-x_{23})} \, \label{R32} \\
{\cal S}_{23} = &\frac{C_{23}}{2\eta_3}\,E_{23,+3} \, = &{\cal R}_{32}^*\,\label{S23}
\end{eqnarray}
where $\eta_1$ is the imaginary part of $\zeta^+_1$, $\eta_3$ is the imaginary part of $\zeta^+_3$, $\gamma_{12}$ and $\gamma_{23}$ are real phases and $x_{12}$ and $x_{23}$ are real spatial positions. Then from the results at the end of Appendix A and equations \eqref{2.24} and \eqref{2.26}, we have
\be\label{Q12}
Q_{12} = (J_1-J_2)\,\frac{2i\eta_1{\cal R}_{12}}{D} \left(1 + 
   \frac{\zeta^-_3-\zeta^+_1}{\zeta^+_3-\zeta^+_1} \,\left|{\cal R}_{32}\right|^2 \right)\,,
\ee
\be\label{Q13}
Q_{13} = (J_1-J_3)\,\frac{4i\eta_1\eta_3{\cal R}_{32}^* {\cal R}_{12}}{D \left(\zeta^+_3-\zeta^+_1 \right)}\,,
\ee
\be\label{Q23}
Q_{23} = -(J_2-J_3)\,\frac{2i\eta_3{\cal R}_{32}^*}{D} \left(1 + 
   \frac{\zeta^+_3-\zeta^-_1}{\zeta^+_3-\zeta^+_1} \,\left|{\cal R}_{12}\right|^2 \right)\,,
\ee
where 
\be\label{Dem}
D = \left(1 + \left|{\cal R}_{12}\right|^2\right)\,\left(1 + \left|{\cal R}_{32}\right|^2\right)
   + \frac{4\eta_1\eta_3}{\left|{\zeta^+_3-\zeta^+_1}\right|^2}\,\left|{\cal R}_{12}\right|^2
   \left|{\cal R}_{32}\right|^2\,,
\ee
and the other three components of $Q$ follow from the assumed symmetry. This solution was originally given in Refs. \cite{ZM,Pub020}, although in a different form.

Contained in these solutions are the two simple $(1,0)$- and $(0,1)$-soliton solutions. For the $(1,0)$-soliton solution, we would have $S^+_{33}$ without a zero and the residue ${\cal R}_{32}$ would be absent from \eqref{Q12}-\eqref{Dem}. In this case only $Q_{12}$ would be nonzero and it would have the form
\be\label{Q1201}
Q_{12} = (J_1-J_2)\,\frac{2i\eta_1{\cal R}_{12}}{1 + \left|{\cal R}_{12}\right|^2} \,,
\ee
which is proportional to a simple $sl(2)$ soliton located at $x=x_{12}$.  Similarly for the $(0,1)$-soliton solution, we would have $R^+_{11}$ without a zero and the residue ${\cal R}_{12}$ would be absent from \eqref{Q12}-\eqref{Dem}. Now only $Q_{23}$ would be nonzero and it would be given by 
\be\label{Q23a}
Q_{23} = -(J_2-J_3)\,\frac{2i\eta_3{\cal R}_{23}^*}{1 + \left|{\cal R}_{32}\right|^2} \,,
\ee
which is also proportional to a simple $sl(2)$ soliton located at $x=x_{23}$. 

Note that there are only two positions and two phases in \eqref{R12} - \eqref{S23} while there are three waves. Thus one wave can be considered to be determined or driven by the other two.  If we take $Q_{13}$ to be driven, then its phase and position is determined by $Q_{12}$ and $Q_{23}$. As to the variety of the solutions that one could obtain here, one finds both analytical and numerical solutions in the literature (see, for instance, \cite{App04}, \cite{App09}, \cite{App06}, \cite{Montes}, \cite{Conforti}, \cite{Fokas}).


Before leaving this solution, there are features of it that we should point out, particularly its asymptotics. To do this, we shall take $|\zeta^+_3 - \zeta^+_1|$ to be generally nonzero and on the order of the imaginary parts or even larger, and the differences $J_1-J_2$ and $J_2-J_3$ to be roughly the same. Let us translate our $x$ coordinate so that one of the two positions will be at zero.  Then the general nature of this solution can be obtained by letting the other position vary from $-\infty$ to $+\infty$. The phases are relatively unimportant in determining this nature so they will be ignored.  Taking $x_{12}=0$, then for $x_{23}$ large, either positive or negative, $Q_{12}$ is an NLS soliton, generally localized around $x=0$. Meanwhile  $Q_{23}$ is also an NLS soliton, but generally localized around $x=x_{23}$, while $Q_{13}$ is generally exponentially small. As $|x_{23}|$ decreases (the two NLS solitons approaching each other) and approaches zero, the two NLS solitons interact and $Q_{13}$ grows. However its growth is limited by two main factors. First, if the real parts, or even the imaginary parts, of the eigenvalues are widely divergent, then the maximum growth in $Q_{13}$ is obviously limited by the $|\zeta^+_3 - \zeta^+_1|$ term in the denominator. Second, the mathematical structure of the solution for $Q_{13}$ is such that the amplitude of $Q_{13}$ will generally be bounded by about a third of that of an NLS soliton with a similar width.  In terms of the 3WRI, the variation of $x_{23}$ from $-\infty$ to $+\infty$ models two NLS solitons (in different waves) coming together, colliding, interacting, and then during the interaction some amount of $Q_{13}$ is produced. As the two NLS solitons pass through each other and then recede, what energy was in $Q_{13}$ is returned to the two NLS solitons, which have suffered no consequences for this collision except for shifts in positions and phases. This regime of the $(1,1)$-soliton is a near-resonant regime wherein only a brief temporary conversion occurs. 

Let us now take up the singular dependence in \eqref{Q12} - \eqref{Dem}, which depends on the difference of the two eigenvalues, which up until now, we have taken to be non-zero. If one takes ${\cal R}_{12}$ and ${\cal R}_{32}$ to be fixed and take the limit where this difference vanishes, then all components of $Q$ vanish. However this ignores the dependence on $x$. As $x$ increases, both ${\cal R}_{12}$ and ${\cal R}_{32}$ will smoothly pass toward zero.  One concludes then that as the difference in the eigenvalues approach zero, for fixed ${C}_{12}$ and ${C}_{32}$, the dominate change to the soliton structure will be that the entire structure will simply shift to larger values of $x$. Alternately, one could achieve the same effect by allowing the amplitudes of ${C}_{12}$ and ${C}_{32}$ to become smaller and smaller, and at such a rate that the general position of the soliton structure would not shift to larger $x$ values. However there is a problem if we allow ${C}_{12}$ and ${C}_{32}$ to vanish: we lose the soliton phase and position information. So there is more involved here than simply a scaling.  

To understand this better, for the sake of argument, let us consider the case of compact support where all components of the scattering matrices can be extended into the entire complex plane. Then from \eqref{RasS}, we have
\begin{equation} \label{R11}
S^+_{33} = R^+_{11}\,R_{22} - R_{12}R_{21}\,.
\end{equation}
Now, $R^+_{11}(\zeta^+_1)=0$ and $S^+_{33}(\zeta^+_3)=0$.  So if $\zeta^+_3\ra \zeta^+_1$, then the left hand side vanishes as well as the first term on the right.  Thus it follows that if $\zeta^+_3 =\zeta^+_1$, then the product $R_{12}R_{21}$ must also vanish (whenever the potential is on compact support). Now, compact support is crucial to this argument and pure soliton solutions are never on compact support. Nevertheless this still indicates that care will have to be used whenever these two eigenvalues approach each other. We will resolve this in more detail in the next section. 

\section{The resonant soliton case}
With the above in mind, let us study this particular singular case where these two transmission coefficients have exactly equal zeros. As one may easily verify, when there is a zero of $R_{11}^+(\z)$ in the UHP, $\z^+_{11,k}$, which exactly matches one of the zeros of $S_{33}^+(\z)$ in the UHP, $\z^+_{33,\ell}$, or when there is a zero of $R_{33}^-(\z)$ in the LHP, $\z^-_{33,k}$, which exactly matches one of the zeros of $S_{11}^-(\z)$ in the LHP, $\z^-_{11,\ell}$, then one will find a singularity in the LDRs, \eqref{2.13eqns} - \eqref{2.16eqns}.  For example, let us take $\z^+_{11,k}$ to be exactly equal to $\z^+_{33,\ell}$ for some value of $k$ and $\ell$.  Then to solve these LDRs, we need the value of $\Theta^+_2(\z_{33,\ell}^+)$ to insert into \eqref{Thm3f}.  This value will follow from \eqref{Thp2f}. However, when we evaluate this equation at $\zeta = \z_{33,\ell}^+$, we see that there will be a term of the form 
$$
{1\over \z^+_{11,k}-\z_{33,\ell}^+}\,\, {R_{12,k}\over R^{+\prime}_{11,k}}\,
   \Theta^+_{1}({\z^+_{11,k}}) E_{12}(\z_{11,k}^+)\,,
$$
which has a denominator that, if $R_{12,k} \Theta^+_{1}({\z^+_{11,k}})$ is nonzero, becomes singular if $\z^+_{11,k}$ is exactly equal to $\z_{33,\ell}^+$. Thus if a solution is to exist, $R_{12,k} \Theta^+_{1}({\z^+_{11,k}})$ must vanish.  That is essentially what happens, which we shall now detail.  For simplicity, we shall assume compact support so that $R$ and $S$, and the Jost solutions $\Phi e^{-i\zeta Jx}$ and $\Psi e^{-i\zeta Jx}$ are entire functions. Later on in this section, we shall briefly discuss how to handle the noncompact support case. 

First, we will collect the consequences of having such paired eigenvalues and introduce our notation.  Since $R_{22}$ and $S_{22}$ are not transmission coefficients, we have the subscript "$2$" free to use.  So let us designate this common zero instead by $\zeta^+_{2,k}$ where the range of $k$ will be over all such zeros in the UHP.  We shall assume this zero is simple in both transmission coefficients. Consider the $(1,1)$ component of \eqref{RasS} at $\zeta = \zeta^+_{2,k}$. Then it follows that the product $S_{23}S_{32}$ must also have a zero at $\zeta = \zeta^+_{2,k}$. In order to avoid double zeros in this product, we shall also assume that $S_{22}S^{+\prime}_{33}\ne R^{+\prime}_{11}$ at $\zeta^+_{2,k}$. Due to compact support, either $S_{23}$ or $S_{32}$ must contain this simple zero.  If we take $S_{23}$ to have this zero, it follows from \eqref{RasS} that $R_{21}$ must also have this same zero.  On the other hand, if we take $S_{32}$ to have this zero, then it similarly follows that $R_{12}$ must also have the same zero. Thus we find that we have a bifurcation with two different possible options. 

Similarly in the LHP, assuming no symmetry, when $\zeta = \zeta^-_{2,k}$ is a simple common zero of both $R^-_{33}$ and $S^-_{11}$ in the LHP, then it follows that $S_{12} S_{21}$ must also have the same simple common zero. Taking $S_{21}$ to be non-zero at $\zeta^-_{2,k}$, we find that  $S_{12} = 0 = R_{32}$ at $\zeta^-_{2,k}$. If we take $S_{12}$ to be non-zero, then $S_{21} = 0 = R_{23}$ at $\zeta^-_{2,k}$.  

These conditions follow directly from the matrix structure of the $sl(3)$ group and compact support. All that has to happen is for a zero of $R^+_{11}(\zeta)$ to match some zero of $S^+_{33}(\zeta)$ in the UHP, with both zeros being simple. Then the other zeros follow, and similarly in the LHP.

\subsection{Bound state structure - UHP} 

In the following, we shall assume no symmetry on the matrix $Q$, which we shall take to be on compact support. Let us turn to the structure of the Jost solutions when $R^+_{11}$ and $S^+_{33}$ have a common zero. In the following, we shall freely make use of \eqref{RasS} and the two orthogonality relations (in tensor notation)
\be\label{Orthog}
 R_{\alpha,1}S_{1,\beta}+ R_{\alpha,2}S_{2,\beta}+ R_{\alpha,3}S_{3,\beta} = \delta^\alpha_\beta = 
   S_{\alpha,1}R_{1,\beta}+ S_{\alpha,2}R_{2,\beta}+ S_{\alpha,3}R_{3,\beta} \,,
\ee
where $\delta^\alpha_\beta$ is the Kronecker delta. 

First, from our definitions of $\mu^\pm$ in \eqref{3.5x}, we have that $\mu_1^\pm$, $R^+_{11}\,\mu_2^+ = \chi_2^+$, $R^-_{33}\mu_2^- = \chi_2^-$ and $\mu_3^\pm$, are entire functions of $\zeta$, where $\chi_2^\pm$ can be taken to be defined by the above relations.  In general $\mu^\pm$ is a meromorphic function while $\chi^\pm$ is an analytic function \cite{Pub249}. Consider the $T$-matrix and its inverse given in Ref. \cite{Pub249}
\be\label{LR}
\mu^+ = \mu^- \cdot T\,, \quad ~\mu^- = \mu^+\cdot T^{-1}\,,
\ee
where
\be \label{LR2}
T = \lb \begin{matrix}
\dfrac{1}{S_{11}^-} & -\dfrac{R_{12}}{S^-_{11}R^+_{11}} & \dfrac{S_{13}}{S^-_{11}} \\
\\
- \dfrac{S_{21}}{S^-_{11}} &  1+\dfrac{R_{12}S_{21}}{S^-_{11}R^+_{11}} & -\dfrac{R_{23}}{S^-_{11}} \\
\\
 \dfrac{R_{31}}{R^-_{33}} & -\dfrac{S_{32}}{R^+_{11}R^-_{33}}  & \dfrac{1}{R^-_{33}}
\end{matrix} \rb ,
\quad
T^{-1} = \lb \begin{matrix}
\dfrac{1}{R^+_{11}} & -\dfrac{S_{12}}{R^+_{11}R^-_{33}} & \dfrac{R_{13}}{R^+_{11}} \\
\\
-\dfrac{R_{21}}{S^+_{33}} & 1+\dfrac{R_{32}S_{23}}{S^+_{33}R^-_{33}} & -\dfrac{S_{23}}{S^+_{33}} \\
\\
\dfrac{S_{31}}{S^+_{33}} & -\dfrac{R_{32}}{R^-_{33}S^+_{33}} & \dfrac{1}{S^+_{33}}
\end{matrix} \rb .
\ee
We have that the lhs of \eqref{LR} are meromorphic functions. It then follows that the matrix product on the rhs of these equations must also be meromorphic and must match any poles and residues found on the lhs. Due to the poles in $\mu^\pm_2$, we can have double poles as well as single poles in \eqref{LR}.  From the first column in the second equation in \eqref{LR}, at $\zeta = \zeta^+_{2,k}$, we obtain
\be\label{Tmp1}
\left[ R_{21}\,\chi^+_2\right](\zeta^+_{2,k}) = 0\,\quad \textrm{and}\quad
     \left[ \mu_1^+ - \frac{1}{S^{+\prime}_{33}} \left( R^\prime_{21}\chi^+_2 + R_{21}\chi^{+\prime}_2\right)
     + \frac{ R^{+\prime}_{11}\,S_{31} }{S^{+\prime}_{33}}\,\mu^+_3\right](\zeta^+_{2,k}) = 0\,.
\ee
From the second column and \eqref{Tmp1}, assuming that $S^-_{11}$ does not vanish at $\zeta^+_{2,k}$, we obtain the single equation
\be\label{Tmp2}
\left[ \chi_2^+ + S_{32}\,\mu^+_3\right](\zeta^+_{2,k}) = 0\,,
\ee
and from the third column, we obtain
\be\label{Tmp3}
\left[ S_{23} \chi_2^+ \right](\zeta^+_{2,k}) = 0\, \quad \textrm{and}\quad
   \left[ \frac{R_{13}S_{33}^{+\prime}}{R_{11}^{+\prime} }\mu^+_1 - \frac{1}{R^{+\prime}_{11}}\left(S_{23}^\prime \chi_2^+ + S_{23}\chi_2^{+\prime}\right) + \mu^+_3
   \right] (\zeta^+_{2,k}) = 0 \,.
\ee
From the above conditions, there are two different options in which these conditions may be satisfied. Thus we have a bifurcation where two different solutions are possible at each such common zero, which we now describe. 

\subsection{UHP - Option UA}

For the first solution, let us take the option to have $S_{23}= 0 = R_{21}$ at $\zeta^+_{2,k}$, which we will call Option UA.  Then the first equations in \eqref{Tmp1} and \eqref{Tmp3} are trivially satisfied while the second ones, along with \eqref{Tmp2}, reduce to 
\be\label{Tmp1a}
   \left[ R_{12}\,\mu^+_1 + \chi_2^+ \right](\zeta^+_{2,k}) = 0\,,\quad 
    \left[S_{13}\mu^+_1 - \,\mu^+_3\right] (\zeta^+_{2,k}) = 0\,,
\ee
where $\mu^+_2$ has a simple pole at $\zeta = \zeta^+_{2,k}$, whose residue is $\chi_2^+(\zeta^+_{2,k})\,/\, R^{+\prime}_{11,k}$. For this option, there is only one independent Jost solution at $\zeta^+_{2,k}$. (For the adjoint Jost functions, there are two.)

\subsection{UHP - Option UB}

For the other solution, the remaining option is to take $S_{32} = 0 = R_{12}$ at $\zeta^+_{2,k}$. Then \eqref{Tmp2} and the first equations in \eqref{Tmp1} and \eqref{Tmp3} give that $\chi^+_2$ must vanish at $\zeta^+_{2,k}$. Thus
\be\label{Tmp2b}
\chi^+_2(\zeta^+_{2,k}) = 0 \,,
\ee
while the second equations in \eqref{Tmp1} and \eqref{Tmp3} reduce to only one condition
\be\label{Opt2b}
 \frac{R_{13,k}}{R_{11,k}^{+\prime} }\mu^+_1(\zeta^+_{2,k}) 
   - \frac{S_{23,k}}{S_{33,k}^{+\prime}}\mu_2^+(\zeta^+_{2,k})
   +  \frac{1}{S_{33,k}^{+\prime}}\mu^+_3(\zeta^+_{2,k})  = 0 \,,
\ee
where by l'Hopital's rule we have made use of $\mu^+_2(\zeta^+_{2,k}) = \chi^{+\prime}_2(\zeta^+_{2,k})\,/\,R^{+\prime}_{11,k}$.  For this option, we have two linearly independent Jost solutions at $\zeta^+_{2,k}$ (while the adjoint only has one).
%

\subsection{Bound state structure - LHP} 

Now let us turn to the LHP and the structure of the Jost solutions when $S^-_{11}$ and $R^-_{33}$ have a common zero. Continuing as before but with the first equation in \eqref{LR} instead, from the first column we obtain
\be\label{Tp1}
\left[ S_{21}\,\chi^-_2\right](\zeta^-_{2,k}) = 0\,,\quad 
     \left[ \mu_1^- - \frac{1}{R^{-\prime}_{33}} \left( S^\prime_{21}\chi^-_2 + S_{21}\chi^{-\prime}_2\right)
     + \frac{ S^{-\prime}_{11}\,R_{31} }{R^{-\prime}_{33}}\,\mu^-_3\right](\zeta^-_{2,k}) = 0\,,
\ee
from the second column and \eqref{Tp1}, assuming that $R^+_{11}$ does not vanish at $\zeta^-_{2,k}$, we obtain
\be\label{Tp2}
\left[ \chi_2^- + R_{32}\,\mu^-_3\right](\zeta^-_{2,k}) = 0\,,
\ee
and from the third column, we obtain
\be\label{Tp3}
\left[ R_{23} \chi_2^- \right](\zeta^-_{2,k}) = 0\,, \quad 
   \left[ \frac{S_{13}R_{33}^{-\prime}}{S_{11}^{-\prime} }\mu^-_1 - \frac{1}{S^{-\prime}_{11}}\left(R_{23}^\prime \chi_2^- + R_{23}\chi_2^{-\prime}\right) + \mu^-_3 \right] (\zeta^-_{2,k}) = 0 \,.
\ee

\subsection{LHP - Option LA}

Continuing as before, under the option $S_{12,k} = 0 = R_{32,k}$ at $\zeta^-_{2,k}$, we have that $\chi^-_2(\zeta^-_{2,k})$ must vanish.  From l'Hopital's rule it follows that $\mu^-_2(\zeta^-_{2,k}) = \chi^{-\prime}_2(\zeta^-_{2,k})\,/\,R^{-\prime}_{33,k}$ giving $\mu^-_2$ to be analytic at $\zeta^-_{2,k}$.  The other equations give only the one  condition
\be\label{Optm1}
\frac{1}{S^{-\prime}_{11,k}}\mu_1^-(\zeta^-_{2,k}) 
   - \frac{S_{21,k}}{S^{-\prime}_{11,k}} \mu^-_2(\zeta^-_{2,k})
     + \frac{ \,R_{31,k} }{R^{-\prime}_{33,k}}\,\mu^-_3(\zeta^-_{2,k}) = 0\,.
\ee
and there are two linearly independent Jost solutions for this option in the LHP (with the adjoint having only one such solution). 

\subsection{LHP - Option LB}

Under the option $S_{21,k} = 0 = R_{23,k}$ at $\zeta^-_{2,k}$, we find that there is only one independent solution for this option.  The other two Jost solutions are given by
\be\label{Optm2}
\chi_2^-(\zeta^-_{2,k}) + R_{32,k}\,\mu^-_3(\zeta^-_{2,k}) = 0\,, \quad 
   \mu^-_1(\zeta^-_{2,k}) - S_{31,k}\,\mu^-_3(\zeta^-_{2,k})=0 \,,
\ee
Here we have again that $\mu^-_2$ has a simple pole at $\zeta^-_{2,k}$, whose residue is $\chi_2^-(\zeta^-_{2,k})\,/\, R^{-\prime}_{33,k}$.

\subsection{LDRs with Resonant Solitons} 

Given the above, one may extend the LDRs given in Ref. \cite{Pub249} to include resonant solitons.  To do this, one would start with Eqs. (3.08)-(3.11) from that reference, which are still correct, even in the case of common zeros between the transmission coefficients. Starting from there, in order to reduce those Eqs. (3.08)-(3.11) to a minimal set of linear independent relations, it is necessary to remove all bound state $\Phi$-type Jost solutions in favor of the $\Psi$-type Jost solutions. As was done for non-common zeros in Ref. \cite{Pub249}, one can extend those equations by using the above options to eliminate the bound state $\Phi$-type Jost solutions. The only change to the previous form of the LDRs, will be additional sums over any common zeros. (As we shall point out later, by treating the common zero case as a limit of different zeros approaching each other in Eqs. (3.15)-(3.19) of Ref. \cite{Pub249}, and by including the four possible options, one can obtain the same results as given below.) 

A priori, in the common zero case and without any symmetry on $Q$, one must allow all four options to coexist. To express this, we need to extend the notation in Ref. \cite{Pub249}. We shall use $\zeta^\pm_{2a,k}$ for the common zeros for option UA and LA with $k = 1,2,\ldots,N^\pm_{2a}$ where $N^\pm_{2a}$ is the number of these common zeros. Similarly $\zeta^\pm_{2b,k}$ will designate the common zeros of options UB and LB where $k = 1,2,\ldots,N^\pm_{2b}$ with $N^\pm_{2b}$ being the number of these common zeros. Meanwhile the current notation used in \eqref{Thp1f}-\eqref{Thm3f} will continue to be used for those zeros which are not common with any in the other transmission coefficient. 

Using the above options, it is straight forward to give these additional sums. One obtains for $\Im{(\z)} \ge 0$,
\be\notag
\Theta^+_1(\z) = \lb \begin{matrix} 1 \\ 0 \\ 0 \end{matrix} \rb
- {1\over 2\pi i} \int_{\cal R}{d\z' \over \z'-\z} \left[ {S_{21}\over S_{11}^-}
   \left( \Theta^-_2( \z^\prime ) E_{21}( \z^\prime )
   + {R_{32} \over R^-_{33}} \Theta_3^-( \z^\prime ) E_{31}( \z^\prime ) \right)
    + {S_{31}\over S_{11}^-}\Theta^-_3( \z^\prime ) E_{31}( \z^\prime )\right]
\ee\be\notag
   - \sum_{k = 1}^{N_{11}^-} \frac{1}{(\z_{11,k}^- - \z)}\, {S_{21,k} \over S_{11,k}^{-\prime}}\,
   \Theta^-_2(\z_{11,k}^-)E_{21}(\z_{11,k}^-)
\ee\be\notag
   - \sum_{k = 1}^{N_{2a}^-} \frac{1}{(\z_{2a,k}^- - \z)}\,
   \left[ \frac{S_{21,k}}{S^{-\prime}_{11,k}} \Theta^-_2(\zeta^-_{2a,k})E_{21}(\zeta^-_{2a,k})
     - \frac{ \,R_{31,k} }{R^{-\prime}_{33,k}}\,\Theta^-_3(\zeta^-_{2a,k})E_{31}(\zeta^-_{2a,k})\right]
\ee\be\label{Thp1ff}
   - \sum_{k = 1}^{N_{2b}^-} \frac{1}{(\z_{2b,k}^- - \z)}\,{S_{31,k}\over S_{11,k}^{-\prime}}\,
     \Theta^-_3(\zeta^-_{2b,k})E_{31}(\zeta^-_{2b,k})\,,
\ee
\be\notag
\Theta^+_2(\z) = \lb \begin{matrix} 0 \\ 1 \\ 0 \end{matrix} \rb
 - {1\over 2\pi i} \int_{\cal R}{d\z' \over \z'-\z} \left[ {R_{12}\over R_{11}^+}\Theta_1^+
   E_{12}(\z^\prime) - {R_{32}\over R_{33}^-}\Theta^-_3 E_{32}(\z^\prime)\right] \notag
\ee
\be\notag
   + \sum_{k=1}^{N^+_{11}} {1\over \z^+_{11,k}-\z}\,\, {R_{12,k}\over R^{+\prime}_{11,k}}\,
   \Theta^+_{1}({\z^+_{11,k}}) E_{12}({\z^+_{11,k}})\,
   +\sum_{k = 1}^{N_{33}^-} \frac{1}{(\z_{33,k}^- - \z)} \, {R_{32,k} \over R_{33,k}^{-\prime}}\,
   \Theta^-_3(\z_{33,k}^-) E_{32}({\z^-_{33,k}})
\ee
\be\label{Thp2ff}
   + \sum_{k=1}^{N^+_{2a}} {1\over \z^+_{2a,k}-\z}\,\, {R_{12,k}\over R^{+\prime}_{11,k}}\,
   \Theta^+_{1}({\z^+_{2a,k}}) E_{12}({\z^+_{2a,k}})\,
   +\sum_{k = 1}^{N_{2b}^-} \frac{1}{(\z_{2b,k}^- - \z)} \, {R_{32,k} \over R_{33,k}^{-\prime}}\,
   \Theta^-_3(\z_{33,k}^-) E_{32}({\z^-_{2b,k}})\,,
\ee
and for $\Im{(\z)} \le 0$,
\be\notag
\Theta^-_2(\z) = \lb \begin{matrix} 0 \\ 1 \\ 0 \end{matrix} \rb
 - {1\over 2\pi i} \int_{\cal R}{d\z' \over \z'-\z} \left[ {R_{12}\over R_{11}^+}\Theta_1^+
   E_{12}(\z^\prime) - {R_{32}\over R_{33}^-}\Theta^-_3 E_{32}(\z^\prime)\right]\notag
\ee
\be\notag
   + \sum_{k=1}^{N^+_{11}} {1\over \z^+_{11,k}-\z}\,\, {R_{12,k}\over R^{+\prime}_{11,k}}\,
   \Theta^+_{1}({\z^+_{11,k}}) E_{12}({\z^+_{11,k}})\,\,
   +\sum_{k = 1}^{N_{33}^-} \frac{1}{(\z_{33,k}^- - \z)} \, {R_{32,k} \over R_{33,k}^{-\prime}}\,
   \Theta^-_3(\z_{33,k}^-)  E_{32}({\z^-_{33,k}})
\ee
\be\label{Thm2ff}
   + \sum_{k=1}^{N^+_{2a}} {1\over \z^+_{2a,k}-\z}\,\, {R_{12,k}\over R^{+\prime}_{11,k}}\,
   \Theta^+_{1}({\z^+_{2a,k}}) E_{12}({\z^+_{2a,k}})\,\,
   +\sum_{k = 1}^{N_{2b}^-} \frac{1}{(\z_{2b,k}^- - \z)} \, {R_{32,k} \over R_{33,k}^{-\prime}}\,
   \Theta^-_3(\z_{33,k}^-) E_{32}({\z^-_{2b,k}})\,,
\ee
\be\notag
\Theta_3^- = \lb \begin{matrix} 0 \\ 0 \\ 1 \end{matrix} \rb
   + {1\over 2\pi i} \int_{\cal R}{d\z' \over \z'-\z} \left[
   {S_{13}\over S_{33}^+}\Theta^+_1( \z^\prime ) E_{13}( \z^\prime ) +
   {S_{23}\over S_{33}^+}\left( \Theta_2^+( \z^\prime ) E_{23}( \z^\prime )
   + {R_{12}\over R_{11}^+} \Theta_1^+( \z^\prime ) E_{13}( \z^\prime )\right) \right]
\ee
\be\notag
   -\sum_{k = 1}^{N_{33}^+} \frac{1}{(\z_{33,k}^+ - \z)} \, {S_{23,k} \over S_{33,k}^{+\prime}}\,
   \Theta^+_2(\z_{33,k}^+)  E_{23}(\z_{33,k}^+)\,
   -\sum_{k = 1}^{N_{2a}^+} \frac{1}{(\z_{2a,k}^+ - \z)} \, {S_{13,k} \over S_{33,k}^{+\prime}}\,
   \Theta^+_1(\z_{2a,k}^+) E_{13}({\z^+_{2a,k}})\,
\ee
\be\label{Thm3ff}
   +\sum_{k = 1}^{N_{2b}^+} \frac{1}{(\z_{2b,k}^+ - \z)}\,
   \left[ \frac{R_{13,k}}{R_{11,k}^{+\prime} }\Theta^+_1(\zeta^+_{2b,k}) E_{13}({\z^+_{2b,k}})
   - \frac{S_{23,k}}{S_{33,k}^{+\prime}}\Theta_2^+(\zeta^+_{2b,k})E_{23}({\z^+_{2b,k}}) \right]\,.
\ee
These equations will be referred to as the ``LDRs with resonances" or ``extended LDRs" while those of \cite{Pub249} will be referred to as the ``regular LDRs" or just plain ``LDRs".

\subsection{Case of Non-compact Support}

The same relations as above could also be obtained in the case of non-compact support. We shall briefly outline here the approach that one would use.  Starting from the relation (2.31) in Ref. \cite{Pub249} and the definition of the $\chi$ states (the Fundamental Analytical Solutions, or FAS) in (2.15) - (2.19) of the same reference, one can obtain $\chi^+ \cdot \chi^{A+} = \textrm{Diagonal}\left[{R_{11}^+,\,R_{11}^+\,S_{33}^+,\,S_{33}^+}\right]$ where $\chi^{A\pm}$ is the adjoint FAS. (A similar expression also exists in the LHP.) Now at a non-common zero of either $R_{11}^+$ or $S_{33}^+$ in the UHP, one has that there will be two linearly independent components of $\chi^+$, as described in \cite{Pub249}. However, if we have a common zero of these two transmission coefficients, then we have that $\chi^+ \cdot \chi^{A+}$ becomes a zero matrix, as well as the middle diagonal component becoming a double zero. From these conditions, one can deduce that at a common zero, $\chi^+$ must have either only one or only two linearly independent columns. (At the same time, the adjoint solutions, $\chi^{A\pm}$, correspondingly must have only two or only one linearly independent solutions.)  These two conditions can then be seen to give rise to the above two options. The only difference would be that the coefficients in \eqref{Tmp1a} - \eqref{Opt2b} will become arbitrary coefficients, generally unrelated to the components of $R$ and $S$ (since the off-diagonal components of these matrices do not generally exist off the real axis, in the non-compact support case). Once these options are established based on the required linear dependences and order of the zeros, then the same LDRs would result.

\section{The Resonant $(1,1)$-Soliton Solution from the Extended LDRs}

Under the anti-hermitian symmetry, 
one can obtain a $(1,1)$-soliton solution when there is only one common zero between $R_{11}$ and $S_{33}$. First, take the scattering data to be in the form of Option A.  Here we have $N^+_{2a} = 1 = N^-_{2a}$ and whence $k=1$ only.  Thus the sum can be omitted.  Furthermore the $k$ subscript can also be deleted since it is no longer needed.
 Then Eqs. \eqref{Thp1ff}-\eqref{Thm3ff} become, for $\Im{(\z)} \ge 0$,
\be\label{Thp1rs}
\Theta^+_1(\z) = \lb \begin{matrix} 1 \\ 0 \\ 0 \end{matrix} \rb
   - \frac{1}{(\z_{2a}^- - \z)}\,
   \left[ C_{21} \Theta^-_2(\zeta^-_{2a})E_{21}(\zeta^-_{2a})
     - C_{31}\,\Theta^-_3(\zeta^-_{2a})E_{31}(\zeta^-_{2a})\right]\,,
\ee
\be\label{Thp2rs}
\Theta^+_2(\z) = \lb \begin{matrix} 0 \\ 1 \\ 0 \end{matrix} \rb
   +{C_{12}\over \z^+_{2a}-\z}\, \Theta^+_{1}({\z^+_{2a}}) E_{12}({\z^+_{2a}})\,,
\ee
and for $\Im{(\z)} \le 0$,
\be\label{Thm2rs}
\Theta^-_2(\z) = \lb \begin{matrix} 0 \\ 1 \\ 0 \end{matrix} \rb
   + {C_{12}\over \z^+_{2a}-\z}\,\Theta^+_{1}({\z^+_{2a}}) E_{12}({\z^+_{2a}})\,,
\ee
\be\label{Thm3rs}
\Theta_3^- = \lb \begin{matrix} 0 \\ 0 \\ 1 \end{matrix} \rb
   - \frac{C_{13}}{\z_{2a}^+ - \z} \,\Theta^+_1(\z_{2a}^+) E_{13}({\z^+_{2a}})\,.
\ee
Let us define 
\begin{eqnarray}
{\cal R}_{12} = &\displaystyle{ \frac{C_{12}}{2\eta }}\,E_{12}({\z^+_{2a}}) \, = &
   \, e^{i\gamma_{12}}\,e^{i\zeta^+_{2a}(J_1 - J_2)(x-x_{12})} \, \label{scR12} \,,\\
{\cal S}_{21} = & \displaystyle{ \frac{C_{21}}{2\eta }}\,E_{21}({\z^-_{2a}}) \, = 
   &{\cal R}_{12}^*\,\label{scS21}\,,\\
{\cal R}_{31} = &\displaystyle{ \frac{C_{31}}{2\eta }}\,E_{31}({\z^-_{2a}}) \, = &
   \, e^{-i\gamma_{31}}\,e^{-i\zeta^-_{2a}(J_1 - J_3)(x-x_{31})} \, \label{scR31}\,, \\
{\cal S}_{13} = &\displaystyle{ \frac{C_{13}}{2\eta }}\,E_{13}({\z^+_{2a}}) \, = 
   &{\cal R}_{31}^*\,,\label{scS13}
\end{eqnarray}
From the above, we find the solution of \eqref{Thp1rs} - \eqref{Thm3rs} to be
\begin{eqnarray}
\Theta^+_1(\zeta) = &\lb \begin{matrix} 1 \\ 0 \\ 0 \end{matrix} \rb
   + \displaystyle{ {2\eta\over (\zeta^-_{2a} - \zeta)D_a} }
     \lb \begin{matrix} i(D-1) \\ -{\cal S}_{21} \\ {\cal R}_{31} \end{matrix} \rb\,,\\
\Theta^\pm_2(\zeta) =  &\lb \begin{matrix} 0 \\ 1 \\ 0 \end{matrix} \rb
   + \displaystyle{ {2\eta{\cal R}_{12} \over (\zeta^+_{2a} - \zeta)D_a} }
      \lb \begin{matrix} 1 \\ -i{\cal S}_{21} \\ i{\cal R}_{31} \end{matrix} \rb\,,\\
\Theta^-_3(\zeta) = &\lb \begin{matrix} 1 \\ 0 \\ 0 \end{matrix} \rb
   + \displaystyle{ {2\eta {\cal S}_{13} \over (\zeta^+_{2a} - \zeta)D_a} }
     \lb \begin{matrix} -1 \\ i{\cal S}_{21} \\ -i{\cal R}_{31} \end{matrix} \rb\,.
\end{eqnarray}
where $\eta$ is the imaginary part of $\zeta^+_{2a}$ and
\be \label{defDa}
D_a = 1 + \left|{\cal R}_{12}\right|^2 + \left| {\cal R}_{31}\right|^2\,.
\ee
From \eqref{2.26} we then obtain the components of the potential matrix 
\begin{eqnarray}
Q_{12} = & i (J_1-J_2) 2\eta \displaystyle{ {{\cal R}_{12}\over D_a } }\,,\label{A12}\\
Q_{13} = & - i (J_1-J_3) 2\eta \displaystyle{ {{\cal S}_{13}\over D_a } }\,,\label{A13}\\
Q_{23} = & - i (J_2-J_3) 2\eta \displaystyle{ {{\cal S}_{13}{\cal S}_{21}\over D_a } } \label{A23}\,,
\end{eqnarray}
with the other three components following from the symmetry relation on the potential matrix. Again, we have only two positions and phases to freely specify.  The position and phase of $Q_{23}$ can be considered to be driven by the other two. 

In the case of Option B we get essentially the same structure but with some indices interchanged. Taking 
\begin{eqnarray}
{\cal R}_{13} = &\displaystyle{ \frac{C_{13}}{2\eta} }\,E_{13}({\z^+_{2b}}) \, = &
   \, e^{i\gamma_{13}}\,e^{i\zeta^+_{2b}(J_1 - J_3)(x-x_{13})} \, \label{scR13} \,,\\
{\cal S}_{31} = & \displaystyle{ \frac{C_{31}}{2\eta } }\,E_{31}({\z^-_{2b}}) \, = 
   &{\cal R}_{13}^*\,\label{scS31}\,,\\
{\cal R}_{32} = &\displaystyle{ \frac{C_{32}}{2\eta} }\,E_{32}({\z^-_{2b}}) \, = &
   \, e^{-i\gamma_{32}}\,e^{-i\zeta^-_{2b}(J_2 - J_3)(x-x_{32})} \, \label{scR32}\,, \\
{\cal S}_{23} = &\displaystyle{ \frac{C_{23}}{2\eta} }\,E_{23}({\z^+_{2b}}) \, = 
   &{\cal R}_{32}^*\,,\label{scS23}
\end{eqnarray}
one obtains for the potential components
\begin{eqnarray}
Q_{12} = & i (J_1-J_2) 2\eta \displaystyle{ {{\cal R}_{13}{\cal R}_{32}\over D_b } }\,,\label{B12}\\
Q_{13} = & i (J_1-J_3) 2\eta \displaystyle{ {{\cal R}_{13}\over D_b } }\,,\label{B13}\\
Q_{23} = & - i (J_2-J_3) 2\eta \displaystyle{ {{\cal S}_{23}\over D_b } } \label{B23}\,,
\end{eqnarray}
where
\be\label{defDemb}
D_b = 1 + \left|{\cal R}_{13}\right|^2 + \left| {\cal R}_{32}\right|^2\,.
\ee
Again we have one component being driven by the other two. In this case, it is $Q_{12}$ which can be considered to be driven. 

The asymptotics of these equations are much easier to handle than those of \eqref{R12} - \eqref{Dem}. In Option A, for $x_{31} << x_{12}$, we have two NLS solitons present, with the one in $Q_{23}$ far to the left and the one in $Q_{12}$ far to the right, with $Q_{13}$ vanishingly small.  As we reverse this and take $x_{12} << x_{31}$, then we find that the original two NLS solitons have vanished and a single NLS soliton has appeared in $Q_{13}$.  

For Option B, if we take $x_{13} << x_{32}$, then we have two NLS solitons present, with the one in $Q_{12}$ far to the left and the one in $Q_{23}$ far to the right. As we reverse this and take $x_{32} << x_{13}$, then we find that the original two NLS solitons have vanished and a single NLS soliton has appeared in $Q_{13}$.  From these asymptotics, one sees clearly that these two options are distinctly different solutions since the spatial orientation of $Q_{12}$ and $Q_{23}$ become reversed. 

There is one case where the solutions of these two options do overlap.  This is when ${\cal R}_{12} = 0 = {\cal S}_{23}$, giving that $Q_{12} = 0 = Q_{23}$ with only $Q_{13}$ remaining nonzero, for each option. This case is also the third case wherein a single $sl(2)$ soliton can occur in this system. As was mentioned earlier, the $(0,1)$-soliton and the $(1,0)$-soliton are NLS solitons, with one in $Q_{12}$ and the other with one in $Q_{23}$. Each one of these NLS solitons only require one zero in one transmission coefficient.  Here we see that $Q_{13}$ can also contain a single NLS soliton, but to obtain this $sl(2)$ soliton, one has to insist that there be a common zero in two {\it different} transmission coefficients, a more difficult objective to achieve practically.  

From the above, one can also argue the instability of the resonant soliton.  The NLS solitons in the $(0,1)$-soliton and the $(1,0)$-soliton solutions are known to be stable. Small variations in any part of their scattering data gives small variations to the NLS solitons.  On the other hand, the existence of the $Q_{13}$ NLS soliton is predicated on more than one coefficient being exactly zero.  Let either one of these coefficients become shifted, then one has the case where the two common eigenvalues are no longer exactly equal. As discussed in Section 2.4 and as seen from \eqref{R12} - \eqref{Dem}, the asymptotics will be distinctly different from those found above for the resonant soliton.  

\section{The Resonate Soliton as a Limit of One Eigenvalue Approaching the Other}

Here we will briefly discuss how the resonate soliton can be obtained as a limit of a zero of $R^+_{11}(\zeta)$ and a zero of $S^+_{33}(\zeta)$ approaching each another. As we have already seen, one has two choices as to how this can be done. Once that is realized, then one must choose one of two options and it becomes straight forward to proceed in that manner and obtain the extended LDRs from the regular LDRs. We will leave that as an exercise to the reader.  

The same thing can be done with the general (1,1)-soliton solution, \eqref{Q12} - \eqref{Dem}. Looking at its structure, one sees if one chooses $C_{12}$ to be proportional to $(\zeta_1^+ - \zeta_3^+)$ in \eqref{R12} and also renormalizes $x_{12}$ to take this into account, then as the eigenvalues approach each other, the solution degenerates into the Option A case.  Similarly by taking $C_{23}$ to be proportional to $(\zeta_1^+ - \zeta_3^+)$ in \eqref{S23}, and renormalizing $x_{32}$, then one obtains the solution for Option B. 

\section{Conclusion}

There were two reasons for this study of the soliton solutions of the $3\times 3$ operator. The first was to obtain a better understanding of the resonant soliton and the scattering data which gives rise to it. The second was to gain a better understanding of the possible soliton solutions in the general $n\times n$ eigenvalue problem. As has been shown here, the resonant soliton can be obtained from either the LDRs or the $(M,N)$-soliton solution, by taking the limit of two eigenvalues in different transmission coefficients approaching each other.  The general $(M,N)$-soliton solution for the $3\times 3$ operator has been constructed in Appendix A from the LRDs. The approach taken there was one that could be extended to the $n\times n$ case, upon being given the LRDs for the $n\times n$ case. (For the case of the $n\times n$ operator, one would include more algebraic equations of the form given in \eqref{Theta1_simple}-\eqref{Theta3_simple}, which could then be solved in the same manner outlined in Appendix A.) Of course, as $n$ increases \cite{ZKSF}, the number of transmission coefficients will also increase, resulting in even more complex forms of resonant soliton solutions, as well as increased options for the number of independent Jost functions at any given resonant eigenvalue.    

The simpler soliton solutions of the $3\times 3$ eigenvalue problem have been presented, including the resonant soliton. In the general case, there are four different transmission coefficients, two in the UHP and two in the LHP. The zeros of these give rise to general $sl(3)$ solitons and generally will express themselves as multiple $sl(2)$ solitons in the appropriate components of the potential matrix $Q$.  However whenever any of these solitons are near to one in the other wave, the corner components of $Q$ could also be excited.  

There are several symmetries which could be applied to the $3\times 3$ potential matrix and each one would generally have different soliton structures \cite{Pub020,ZM}.  For the anti-hermitian symmetry, the eigenvalues and normalization coefficients in the LHP become equal to the complex conjugates of their counterparts in the UHP. In this case, one can designate the $sl(3)$ soliton solutions by $(M,N)$ where $M$ and $N$ are the number of zeros in the two transmission coefficients in the UHP. The simple case of a $(1,0)$- or $(0,1)$-soliton solution is identical to an NLS soliton.  These solitons will only be found in one of the two off-diagonal components which are directly next to the diagonal of $Q$.  Moving up to the next case of a $(1,1)$-soliton solution, as long as the two transmission coefficients do not have eigenvalues close to one another and as long as the NLS solitons do not have their positions close together, the $(1,1)$-soliton solution is essentially a sum of two $(1,0)$- and $(0,1)$-solitons. 


When the magnitude of the difference between the two eigenvalues become significantly smaller than their imaginary parts, the $sl(3)$ $(1,1)$-soliton approaches a (nonlinear) resonance.  In this realm, the $sl(3)$ soliton is found to have a presence in all off-diagonal components of $Q$.  As the positions of the two NLS   solitons come closer together, the amplitude of the $sl(3)$ soliton in the $Q_{13}$ wave increases while the components in the other waves $Q_{12}$ and $Q_{23}$ correspondingly decrease. However the component of the soliton in the $Q_{13}$ wave never fully forms into a NLS soliton. As the difference between the eigenvalues becomes smaller and smaller, the $Q_{13}$ component comes closer and closer to forming a complete NLS soliton. This is the regime of a near nonlinear resonance wherein the soliton in $Q_{13}$ never does fully form. 

When the eigenvalues become exactly equal, one has the fully nonlinear resonant case. Here the $(1,1)$-soliton solutions are best given by \eqref{defDa}-\eqref{A23} and \eqref{B12}-\eqref{defDemb}. However as pointed out earlier, they can also be obtained as a limit process from \eqref{R12} - \eqref{Dem}. The range of solutions here vary from widely separated NLS solitons in $Q_{12}$ and $Q_{23}$ with essentially no component of the $sl(3)$ soliton in $Q_{13}$, to the NLS solitons in $Q_{12}$ and $Q_{23}$ asymptotically being completely absorbed into a fully formed NLS soliton in $Q_{13}$. The asymptotic limit of this latter solution provides us with the third $sl(2)$ soliton.  This asymptotic limit for \eqref{defDa}-\eqref{A23} follows upon taking $C_{12} = 0$ and the same can be obtained from \eqref{B12}-\eqref{defDemb} upon taking $C_{23}=0$. 

As shown earlier, there are two different structures for resonant solitons, depending on whether there is one (Option A) or two (Option B) independent Jost solutions at the eigenvalue. The general solution for Option A of the resonant $(1,1)$-soliton has only one independent Jost solution at the eigenvalue and is given by \eqref{defDa} - \eqref{A23}.  Equations \eqref{B12} - \eqref{defDemb} give the solution for Option B which has two independent Jost solutions at the eigenvalue. As pointed out earlier, the solutions for these two options are distinctly different except for the one common solution where the normalization coefficients $C_{12}$ and $C_{32}$ both vanish. When that happens, the solution degenerates into the form of a single NLS soliton in $Q_{13}$ (and $Q_{31}$) with all other components of $Q$ zero.  Here we have a single NLS $1$-soliton solution resulting from {\it two} zeros, each in a different transmission coefficient. 
As mentioned earlier, in this resonant case, viewed from the evolution of the 3WRI, the solution for one option is just the time-reversed solution of the other. 

The reason this resonance is a ``nonlinear" resonance is that it requires an exact equality between two eigenvalues, each of which is a zero of a different transmission coefficient, and each of which is independent of the other. From the direct scattering problem, the positions of these zeros are nonlinear functions of the potential matrix $Q$. From the inverse scattering problem, it is the eigenvalue which determines the spatial shape of the amplitude and phase of a soliton solution.  (The normalization coefficient determines only the position and overall phase of the soliton solution.) When these two eigenvalues are equal, then the shape of the solitons in $Q_{12}$ and $Q_{23}$ are such that, in the language of the 3WRI, as these two waves collide, they will coherently and constructively interfere with each other, transferring all their energy into the third wave, $Q_{13}$. Recall that for plane waves in the 3WRI, the resonance conditions are that the sums of two of the wave-vectors and the sums of two of the frequencies must equal those of the third.  However for plane waves, one never achieves a complete conversion since the inverse processes will limit and prevent complete conversion. Thus the resonant soliton is a solution to the 3WRI problem of how to shape and phase the envelopes of two colliding beams, $Q_{12}$ and $Q_{23}$, such that one could achieve a total conversion into the third wave, $Q_{13}$. The condition for this nonlinear resonance and complete conversion is the equality of the eigenvalues. Similarly, any fully resonant (every eigenvalue for one wave being paired with another in the other wave) $(N,N)$-soliton solution would also do the same.

\appendix

\section{Appendix A: Derivation of the general soliton solution; non-common zeros}
We obtain the $(M,N)$-soliton solution in the absence of any symmetry. We assume no common zeros between the $M$ zeros and the $N$ zeros and that each $M$ zero and each $N$ zero is simple. One can substitute the expressions for the $\Theta_2$'s from \eqref{2.14eqns} and \eqref{2.15eqns} into the remaining expressions \eqref{2.13eqns} and \eqref{2.16eqns} to get
\be 
\begin{aligned}
\Theta^+_1(\z) & = \lb \begin{matrix} 1 \\ 0 \\ 0 \end{matrix} \rb
      - \lb \begin{matrix} 0 \\ 1 \\ 0 \end{matrix} \rb 
      \sum_{k = 1}^{N_{11}^-} \frac{1}{(\z_{11,k}^- - \z)}\, {S_{21,k} \over S_{11,k}^{-\prime}}\,
       E_{21}(\z_{11,k}^-) \\  
   & - \sum_{k = 1}^{N_{11}^-} \frac{1}{(\z_{11,k}^- - \z)}\, {S_{21,k} \over S_{11,k}^{-\prime}}\,
   \sum_{\ell =1}^{N^+_{11}} {1\over \z^+_{11,\ell}-\z_{11,k}^-}\,\, {R_{12,\ell}\over R^{+\prime}_{11,\ell}}\,
   \Theta^+_{1}({\z^+_{11,\ell}}) E_{12}(\z_{11,\ell}^+) E_{21}(\z_{11,k}^-)\\
   &   - \sum_{k = 1}^{N_{11}^-} \frac{1}{(\z_{11,k}^- - \z)}\, {S_{21,k} \over S_{11,k}^{-\prime}}\,
    \sum_{\ell = 1}^{N_{33}^-} \frac{1}{(\z_{33,\ell}^- - \z_{11,k}^-)} \, {R_{32,\ell} \over R_{33,\ell}^{-\prime}}\,
   \Theta^-_3(\z_{33,\ell}^-) E_{32}(\z_{33,\ell}^-) E_{21}(\z_{11,k}^-)
   \,,
   \end{aligned}
\ee
\be 
\begin{aligned}
\Theta_3^- & = \lb \begin{matrix} 0 \\ 0 \\ 1 \end{matrix} \rb
      - \lb \begin{matrix} 0 \\ 1 \\ 0 \end{matrix} \rb \sum_{k = 1}^{N_{33}^+} \frac{1}{(\z_{33,k}^+ - \z)} \, {S_{23,k} \over S_{33,k}^{+\prime}}\,
     E_{23}(\z_{33,k}^+)\,\\
    & -\sum_{k = 1}^{N_{33}^+} \frac{1}{(\z_{33,k}^+ - \z)} \, {S_{23,k} \over S_{33,k}^{+\prime}}\,
   \sum_{\ell =1}^{N^+_{11}} {1\over \z^+_{11,\ell}-\z_{33,k}^+}\,\, {R_{12,\ell}\over R^{+\prime}_{11,\ell}}\,
   \Theta^+_{1}({\z^+_{11,\ell}}) E_{12}(\z_{11,\ell}^+)  E_{23}(\z_{33,k}^+)\,\\
   & -\sum_{k = 1}^{N_{33}^+} \frac{1}{(\z_{33,k}^+ - \z)} \, {S_{23,k} \over S_{33,k}^{+\prime}}\,
   \sum_{\ell = 1}^{N_{33}^-} \frac{1}{(\z_{33,\ell}^- - \z_{33,k}^+)} \, {R_{32,\ell} \over R_{33,\ell}^{-\prime}}\,
   \Theta^-_3(\z_{33,\ell}^-) E_{32}(\z_{33,\ell}^-)  E_{23}(\z_{33,k}^+)\,,
\end{aligned}
\ee
or, in a more compact notation,
\be \label{Theta1_simple}
\Theta_1^+(\z)= \lb \begin{matrix} 1 \\ -\mathcal{F}_2(\z) \\ 0 \end{matrix} \rb
- \sum_{\ell =1}^{N_{11}^+}\mathcal{F}_{1,\ell}(\z) \Theta_1^+(\z_{11,\ell}^+) - \sum_{\ell =1}^{N_{33}^-}\mathcal{F}_{3,\ell}(\z) \Theta_3^-(\z_{33,\ell}^-) \,,
\ee
\be  \label{Theta3_simple}
\Theta_3^-(\z)= \lb \begin{matrix} 0 \\ -\mathcal{G}_2(\z) \\ 1 \end{matrix} \rb
- \sum_{\ell =1}^{N_{11}^+}\mathcal{G}_{1,\ell}(\z) \Theta_1^+(\z_{11,\ell}^+) - \sum_{\ell =1}^{N_{33}^-}\mathcal{G}_{3,\ell}(\z) \Theta_3^-(\z_{33,\ell}^-) \,,
\ee
where the $\mathcal{F}$'s and $\mathcal{G}$'s are defined in the obvious manner. Now, using these expressions and evaluating $\Theta_1^+(\z)$ at $\z = \z_{11,k}^+$,
\be 
\Theta_1^+(\z_{11,k}^+)= \lb \begin{matrix} 1 \\ -\mathcal{F}_2(\z_{11,k}^+) \\ 0 \end{matrix} \rb
- \sum_{\ell =1}^{N_{11}^+}\mathcal{F}_{1,\ell}(\z_{11,k}^+) \Theta_1^+(\z_{11,\ell}^+) - \sum_{\ell =1}^{N_{33}^-}\mathcal{F}_{3,\ell}(\z_{11,k}^+) \Theta_3^-(\z_{33,\ell}^-) \,,
\ee
for $k = 1,2,\dots ,N_{11}^+$, and evaluating $\Theta_3^-(\z)$ at $\z = \z_{33,k}^-$,
\be
\Theta_3^-(\z_{33,k}^-)= \lb \begin{matrix} 0 \\ -\mathcal{G}_2(\z_{33,k}^-) \\ 1 \end{matrix} \rb
- \sum_{\ell =1}^{N_{11}^+}\mathcal{G}_{1,\ell}(\z_{33,k}^-) \Theta_1^+(\z_{11,\ell}^+) - \sum_{\ell =1}^{N_{33}^-}\mathcal{G}_{3,\ell}(\z_{33,k}^-) \Theta_3^-(\z_{33,\ell}^-) \,,
\ee
$k = 1,2,\dots ,N_{33}^-$, which together are a system of $N_{11}^+ + N_{33}^-$ vector equations for the $\Theta_1^+(\z_{11,k}^+)$'s and $\Theta_3^-(\z_{33,k}^-)$'s. Solving this system, and plugging them back into the original expressions \eqref{Theta1_simple} and \eqref{Theta3_simple}, we recover the closed form solutions for $\Theta_1^+(\z)$ and $\Theta_3^-(\z)$, as well as $\Theta_2^{\pm}(\z)$. Explicitly, we may write the system in matrix form, to wit:
\be 
\mathcal{M}\vartheta = \sigma \,,
\ee
where
\be 
\vartheta ^T = \left[ \Theta_1^+(\z_{1}^+),\dots ,\Theta_1^+(\z_{11,N_{11}^+}^+), \Theta_3^-(\z_{33,1}^-),\dots , \Theta_3^-(\z_{33,N_{33}^-}^-) \right] \,,
\ee
\be 
\sigma ^T = \left[ \sigma_1^+ ,\dots , \sigma_{N_{11}^+}^+, \sigma_1^-, \dots \sigma_{N_{33}^-}^-,  \right]  \,,
\ee
\be 
\sigma_k^+ = \lb \begin{matrix} 1 \\ -\mathcal{F}_2(\z_{11,k}^+) \\ 0 \end{matrix} \rb , \quad 
\text{for} ~~ k = 1,2,\dots ,N_{11}^+,\quad
\sigma_k^- = \lb \begin{matrix} 0 \\ -\mathcal{G}_2(\z_{33,k}^-) \\ 1 \end{matrix} \rb , \quad
\text{for} ~~ k = 1,2,\dots ,N_{33}^-\,,
\ee
\be
\mathcal{M} = \lb \begin{matrix}  \mathcal{F}_{1,1}(\z_{1}^+)  +1 & \cdots & \mathcal{F}_{1,N_{11}^+}(\z_{1}^+) & \mathcal{F}_{3,1}(\z_{1}^+)    & \cdots & \mathcal{F}_{3,N_{33}^-}(\z_{1}^+) \\
\mathcal{F}_{1,1}(\z_{11,2}^+)   & \cdots & \mathcal{F}_{1,N_{11}^+}(\z_{11,2}^+) & \mathcal{F}_{3,1}(\z_{11,2}^+)    & \cdots & \mathcal{F}_{3,N_{33}^-}(\z_{11,2}^+) \\
\vdots   & \cdots & \vdots  & \vdots  & \cdots & \vdots \\
\mathcal{F}_{1,1}(\z_{11,N_{11}^+}^+)   & \cdots & \mathcal{F}_{1,N_{11}^+}(\z_{11,N_{11}^+}^+) +1 & \mathcal{F}_{3,1}(\z_{11,N_{11}^+}^+)  &  \cdots & \mathcal{F}_{3,N_{33}^-}(\z_{11,N_{11}^+}^+) \\
 \mathcal{G}_{1,1}(\z_{33,1}^-)    & \cdots & \mathcal{G}_{1,N_{11}^+}(\z_{33,1}^-) & \mathcal{G}_{3,1}(\z_{33,1}^-)  +1 & \cdots & \mathcal{G}_{3,N_{33}^-}(\z_{33,1}^-) \\
\mathcal{G}_{1,1}(\z_{33,2}^-)    & \cdots & \mathcal{G}_{1,N_{11}^+}(\z_{33,2}^-) & \mathcal{G}_{3,1}(\z_{33,2}^-)    & \cdots & \mathcal{G}_{3,N_{33}^-}(\z_{33,2}^-) \\
\vdots   & \cdots & \vdots  & \vdots  & \cdots & \vdots \\
\mathcal{G}_{1,1}(\z_{33,N_{33}^-}^-)  & \cdots & \mathcal{G}_{1,N_{11}^+}(\z_{33,N_{33}^-}^-) & \mathcal{G}_{3,1}(\z_{33,N_{33}^-}^-)  &  \cdots & \mathcal{G}_{3,N_{33}^-}(\z_{33,N_{33}^-}^-) +1
\end{matrix} \rb \,,
\ee
where the $+1$'s occur with the diagonal terms only. Then, (assuming $\mathcal{M}$ is invertible) the general solutions for $\Theta_1^+(\z_{11,k}^+)$ and $\Theta_3^+(\z_{33,k}^-)$ are (by Cramer's rule)
\be 
\Theta_1^+(\z_{11,k}^+) = \frac{\det(\mathcal{M}_k)}{\det(\mathcal{M})} \,, \quad 
\Theta_3^+(\z_{33,k}^-) = \frac{\det(\mathcal{M}_{N_{11}^+ + k})}{\det(\mathcal{M})} \,,
\ee
where $\mathcal{M}_k$ is the matrix formed by replacing the $k$th column of $\mathcal{M}$ with $\sigma$ and $\mathcal{M}_{N_{11}^+ + k}$ is the matrix formed by replacing the $(N_{11}^+ + k)$th column of $\mathcal{M}$ with $\sigma$. Arranging all terms properly, we obtain
\be \label{Theta1_closed}
\Theta_1^+(\z)= \lb \begin{matrix} 1 \\ -\mathcal{F}_2(\z) \\ 0 \end{matrix} \rb
- \sum_{\ell =1}^{N_{11}^+}\mathcal{F}_{1,\ell}(\z) \frac{\det(\mathcal{M}_\ell)}{\det(\mathcal{M})} - \sum_{\ell =1}^{N_{33}^-}\mathcal{F}_{3,\ell}(\z) \frac{\det(\mathcal{M}_{N_{11}^+ + \ell})}{\det(\mathcal{M})} \,,
\ee
\be
\Theta^+_2(\z)  = \lb \begin{matrix} 0 \\ 1 \\ 0 \end{matrix} \rb
   + \sum_{k=1}^{N^+_{11}} {C_{12,k}\over \z^+_{11,k}-\z}\,\,
   \frac{\det(\mathcal{M}_k)}{\det(\mathcal{M})} E_{12}(\z_{11,k}^+)\,
   +\sum_{k = 1}^{N_{33}^-} \frac{C_{32,k}}{(\z_{33,k}^- - \z)} \,   
   \frac{\det(\mathcal{M}_{N_{11}^+ + k})}{\det(\mathcal{M})} E_{32}(\z_{33,k}^-)\,,
\ee
\be
\Theta^-_2(\z) = \Theta^+_2(\z)\,,
\ee
\be  \label{Theta3_closed}
\Theta_3^-(\z)= \lb \begin{matrix} 0 \\ -\mathcal{G}_2(\z) \\ 1 \end{matrix} \rb
- \sum_{\ell =1}^{N_{11}^+}\mathcal{G}_{1,\ell}(\z) \frac{\det(\mathcal{M}_\ell)}{\det(\mathcal{M})} - \sum_{\ell =1}^{N_{33}^-}\mathcal{G}_{3,\ell}(\z) \frac{\det(\mathcal{M}_{N_{11}^+ + \ell})}{\det(\mathcal{M})} \,.
\ee
Under this framework, these results scale up for the soliton solutions of the general $n\times n$ operator \cite{Pub250}.

\end{document}